\documentclass[aps,prl,twocolumn,superscriptaddress,showpacs,citeautoscript]{revtex4-1}

\usepackage{amsmath,amssymb}
\usepackage{bm}
\usepackage{graphicx}
\usepackage{epstopdf}
\usepackage{latexsym}
\usepackage{subfigure}
\usepackage{color}
\usepackage{dsfont} 
\usepackage{wasysym} 

\begin{document}

\title{Ferromagnetism in itinerant two-dimensional t$_{2g}$ systems}

\author{Gang Chen}
\email{email address: gang.chen@colorado.edu}
\affiliation{Department of Physics, University of Colorado, Boulder, CO, 80309-0390, U.S.A.}
\author{Leon Balents}
\affiliation{Kavli Institute for Theoretical Physics, University of
  California, Santa Barbara, CA, 93106-4030, U.S.A.}

\begin{abstract}
  Motivated by the recent indications of ferromagnetism
  in transition metal oxide heterostructures, we propose a
  possible mechanism to generate ferromagnetism for itinerant t$_{2g}$
  systems in two spatial dimensions that does not rely on the coupling
  between local moments and conduction electrons.  We particularly
  emphasize the orbital nature of different bands and show that, when
  the Fermi level lies near the bottom of the upper bands, a
  non-perturbative interaction effect due to the
  quasi-one-dimensional nature of the upper bands may drive a
  transition to a state in which the upper bands are ferromagnetically
  polarized.  In the quasi-one-dimensional limit, the full
  thermodynamics may be obtained exactly.   We discuss the connection between
  our mechanism with several itinerant t$_{2g}$ systems that may have
  ferromagnetic instabilities.
\end{abstract}

\date{\today}

\pacs{71.30.+h, 71.10.Ay, 71.45.Lr, 75.30.Fv}

\maketitle

Possible ferromagnetism at polar interfaces between SrTiO$_3$ (STO)
and other oxides such as LaAlO$_3$ (LAO) or GdTiO$_3$
(GTO)\cite{PhysRevX.2.021014,Brinkman07,PhysRevLett.107.056802,Ariando11,bert2011direct,li2011coexistence} has raised considerable excitement.  Such
ferromagnetism is remarkable as the electrons are believed to reside
nearly completely in the t$_{2g}$ bands of the STO, where there are no
localized partially filled shells to form local
moments. Ferromagnetism in purely
itinerant systems, while envisioned long ago\cite{Stoner33}, is quite rare in practice;
most examples may be at least partially attributed to local moments
formed by partially filled d-shells (and often other delocalized
electrons). 

In this paper, we discuss the possibility of ferromagnetism in t$_{2g}$
systems of this type.  In a typical metallic state, ferromagnetism is
unfavorable because of the kinetic energy cost.  This is believed to be
overcome at very low density, where the dimensionless inter-electron
distance $r_s \gtrsim 30$\cite{PhysRevB.39.5005, PhysRevLett.45.566}),
and also very close to the Mott metal-insulator transition, when the
electron filling is close to an integer.  In the former regime,
ferromagnetism and indeed metallicity as well are extremely fragile to
disorder, and may be disregarded in almost all practical situations.
The latter situation {\em might} be thought to apply to the
aforementioned LAO/STO and GTO/STO interfaces, in which there is an
intrinsic mechanism for high carrier density: the polar
discontinuity\cite{Harrison78}.  LAO and GTO have a structure of polar
(001) layers: La$^{3+}$O$^{2-}$ has a net charge of $+1$ per unit cell,
while Al$^{3+}$(O$^{2-}$)$_2$ has a net charge of $-1$ per unit cell
(the same counting holds for GTO).  STO by contrast is non-polar.  At an
ideal (i.e. without atomic reconstruction or compensating defects)
interface between two such materials, an electron gas is predicted to
arise with a carrier density of {\sl half} an electron per
two-dimensional unit cell.  This translates, using the unit cell of STO,
into a two-dimensional carrier density of $n=3.5 \times 10^{14}
cm^{-2}$, which is extremely large by semiconductor standards.  However,
this still corresponds to a fractional Ti site occupation $x<0.5$, and
probably more properly $x<0.2$, taking into account the spread of the
electrons normal to the
interface.\cite{PhysRevB.86.125121,joshua2011universal,son2009density,delugas2011spontaneous}\
Modern computational studies have put strong restrictions on
ferromagnetism due to Mott physics in Hubbard
models\cite{PhysRevLett.66.369}, the most recent studies arguing it is
absent in the two-dimensional Hubbard model for fractional site
occupation $x\lesssim 0.7$, {\sl even when the on-site Hubbard
  interaction $U\rightarrow
  \infty$}\cite{liu2012phases,carleo2011itinerant}.  STO 2DEGs are well
below this degree of site occupation, so Mott physics cannot be invoked
to explain ferromagnetism.

\begin{figure}[h]
  \begin{center}
  \scalebox{0.9}{\includegraphics[width=\columnwidth]{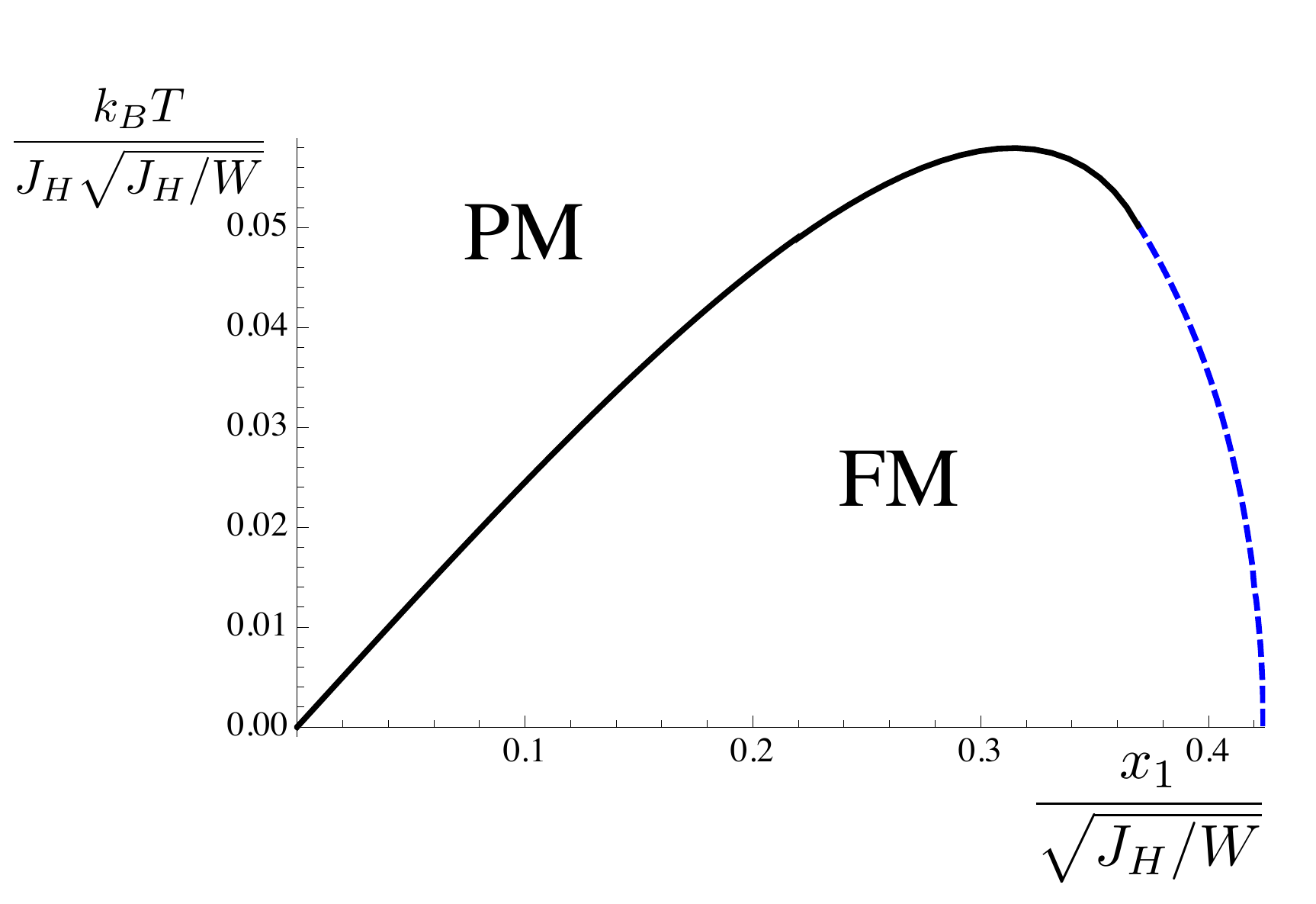}}
  \end{center}
\caption{Phase boundary in the 1d limit, $t'=0$.  Here $x_1$ is the
  occupation per site of the $xz$ or $yz$ orbitals, $J_H$ is the
  renormalized Hund's coupling (see text), $W$ is an energy scale
  of order the hopping, and it is assumed that $x_1, J_H/W \ll 1$.  Solid and dashed
  lines denote continuous and first order transitions, respectively.}
        \label{fig:phase}
\end{figure}

Instead, we propose here that an unusual enhanced tendency to
ferromagnetism may occur due to the quasi-one-dimensionality of
certain bands in these materials, which in turn arises due to the
directionality of the $t_{2g}$ orbitals involved.  The enhanced
tendency to magnetism is a {\em non-perturbative} effect of the
Hubbard $U$ interaction, which leads to strong scattering
in one dimensional subbands with low filling.  The non-perturbative
effects can be controlled by virtue of exact results and bosonization
methods which are particular to one dimensional problems.  Due to the
non-perturbative effect, ferromagnetism is induced by even very weak
atomic Hund's exchange on the Ti atom (see below).   We argue that the
ferromagnetism survives in sufficiently anisotropic two dimensional
systems.  It occurs only for low filling of the $xz$/$yz$ subbands,
where the majority of the polarization resides. The
central result of our calculations is summarized by the phase diagram
in Fig.~\ref{fig:phase}.   

For our discussion, we will require a few particulars of the
conduction band states in STO, which are well-established. The low
lying octahedral t$_{2g}$ crystal field levels of Ti comprise $yz$,
$xz$, and $xy$ orbitals.  Owing to its directionality, hopping $t$ in
the plane of a given orbital is much larger than the hopping $t'$
normal to the plane
(values in the literature are in the range $0.03< t'/t<
0.15$\cite{PhysRevB.86.125121}).  Thus in a bulk system with cubic
symmetry, there are three bands each of which disperses predominantly
in 2 of the 3 cartesian directions.    When confinement is introduced in
the $z$ direction for a (001) interface,
an $xy$ subband is lowest energy and disperses fairly uniformly in the two
dimensional plane, while $xz$ and $yz$ subbands are higher in energy and
approximately {\sl one-dimensional}.

The reduction of the kinetic energy of the $xz$ and $yz$ subbands suggests
we consider them for possible ferromagnetic polarization.  We therefore
adopt a minimal model with three two-dimensional sub-bands for the (001) interface, 
with the Hamiltonian
${\mathcal H}= {\mathcal H}_0 + {\mathcal H}_I$, where the kinetic energy is
\begin{eqnarray}
  \label{eq:1}
  {\mathcal H}_0 &= & \sum_{{\bf k},\alpha} \frac{k_x^2+k_y^2}{2m_0} d_{0\alpha}^\dagger({\bf k})
  d_{0\alpha}^{\vphantom\dagger}({\bf k})      \nonumber
 \\
  && +
  \sum_{{\bf k},\alpha,i=x,y} \left(\Delta+\frac{k_i^2}{2m_i}\right) d_{i\alpha}^\dagger( {\bf k})
  d_{i\alpha}^{\vphantom\dagger}({\bf k}).
\end{eqnarray}
Here $d_{0\alpha},d_{x\alpha},d_{y\alpha}$ describe the $xy$, $xz$, and
$yz$ bands (with spin polarization $\alpha=\uparrow,\downarrow$),
respectively, $m_i$ is an effective mass, and $\Delta$ is the subband crystal field
splitting. We have assumed a tetragonal crystal field symmetry for the (001) interface in
Eq.~\eqref{eq:1}, so $m_x = m_y$.  
  Since we always consider the band bottom, this is equivalent
to taking a tight binding model with hopping amplitude $t_i=1/(2m_i a^2)$,
where $a$ is the lattice spacing.  Here we make an approximation 
$t_0 \approx t_x = t_y \equiv t$ so that all the effective masses are equal. 
We take on-site interactions, of the
form
\begin{eqnarray}
  \label{eq:2}
  {\mathcal H}_I & = & U \sum_{{\bf r} ,i}  n_{i\uparrow}({\bf r}) n_{i\downarrow}({\bf r}) + U'
  \sum_{{\bf r},i\neq j} n_{i}({\bf r}) n_{j}({\bf r})  \nonumber \\
  && - J_H \sum_{{\bf r},i\neq j} {\bf S}_i({\bf r}) \cdot {\bf S}_j({\bf r}),
\end{eqnarray}
where $n_{i\alpha}({\bf r}) = d_{i\alpha}^\dagger({\bf r})
d_{i\alpha}^{\vphantom\dagger}({\bf r})$, $n_i ({\bf r})= \sum_\alpha
n_{i\alpha}({\bf r})$, and ${\bf S}_i ({\bf r})= \frac{1}{2}
\sum_{\alpha\beta} d_{i\alpha}^\dagger( {\bf r} )
\boldsymbol{\sigma}_{\alpha\beta} d_{i\beta}^{\vphantom\dagger}({\bf
  r})$.  As usual, we expect the intra-orbital interaction Hubbard $U$
to be the largest interaction, with the inter-orbital interaction $U'$
and the Hund's coupling $J_H$ rather smaller, $U'/U, J_H/U \lesssim
0.3$.  Other interactions are typically at least an order of magnitude
smaller in 3d transition metal compounds, and interactions between
different Ti sites are strongly screened.  To proceed, we first
project $\mathcal{H}_I$ onto the 2d subbands, which replaces the
couplings by renormalized reduced ones, $U \rightarrow
U/z,U'\rightarrow U'/z, J_H\rightarrow J_H/z$ where $z$ is roughly the
number of STO unit cells over which the subbands are spread (strictly
speaking these factors depend on the bands involved, but below
interactions play a key role only for the $xz/yz$ subbands -- for which $z
\geq 4$ obtains based on subband modeling).  We henceforth
absorb this renormalization into the couplings.

To analyze the effect of interactions, we now treat the $U'$ and $J_H$
as small (which they are, relative to $U$), and consider possible
ferromagnetic instabilities they induce.  With $U'=J_H=0$, the
Hamiltonian is decoupled to three single-band problems, and hence does
not support ferromagnetism.  However, crucially, the $xz$ and $yz$
sub-systems become extremely {\em susceptible} to ferromagnetism when
they contain a low density of electrons.  What we require is the free
energy of each orbital subsystem as a function of its magnetization,
including the effects of strong on-site $U$.  For the $xy$ subband,
which is two-dimensional, an on-site interaction $U$ has little
effect, and in the low density (per lattice site) limit studied here
the interactions can be exactly treated by a standard T-matrix ladder
summation.  The result is simply a Fermi liquid with small Landau
parameters, which can be neglected at the level of the present
consideration.  {\em A posteriori}, it is justified to assume the
magnetization $M_0$ of the $xy$ subband is small, so that we can just
approximate its free energy by the quadratic form
$M_0^2/(2\chi_{2d})$, where $\chi_{2d}$ is the susceptibility for such
a 2d system.  Neglecting the Fermi liquid correction, this is
$\chi_{2d} = ma^2/(2\pi)=1/(4\pi t)$.

For the $xz$ and $yz$ subbands, however, due to their
one-dimensionality, the situation is radically different.  Remarkably,
it is known that the susceptibility, $\chi_{1d}$, of a 1d electron gas
(1DEG) is highly divergent at low
density.\cite{PhysRevB.6.930,schulz1991correlated} In particular, it
actually diverges as $\chi_{1d} \sim 1/(W x_1^2)$, where $W$ is an
energy scale and $x_1$ is the occupation per site, for any non-zero
$U$.  This is a strong interaction effect: the ratio of the
interacting to free fermion susceptibility $\chi_{1d}/\chi_{ff}
\rightarrow\infty$ diverges for $x_1\rightarrow 0$. 

We can explain the enhanced susceptibility, and even obtain a general
result for the free energy versus magnetization, starting from the
fact that low energy scattering is enhanced in one
dimension.  In particular, for an arbitrary repulsive interaction, the
reflection probability for a pair of scattering particles approaches
unity when their energy approaches its minimum -- this is true {\em
  only} in one dimension.  Consequently, the electrons in a low
density 1DEG are almost unable to exchange, and the energy of the
ground state becomes almost independent of spin, and equal to that of
{\em spinless} fermions.  In fact, there is a parametrically weak
residual exchange coupling, which occurs due to the small transmission
probability of colliding electrons.  Because the charge degrees of
freedom are well-ordered on this exchange scale, {\em the spin dependence of the ground
  state energy is exactly that of a one-dimensional Heisenberg
  antiferromagnetic chain} with an effective exchange interaction
$J_{\rm eff}$ much smaller than the 1d Fermi energy $\epsilon_F$, and
one ``site'' per electron in the 1DEG.  We expect $J_{\rm eff}/\epsilon_F$ to
vanish as $x_1\rightarrow 0$, and since $\epsilon_F \sim t x_1^2$, we
guess $J_{\rm eff} \sim W x_1^3$.  We have checked that this agrees
with all known exact results for the susceptibility of the 1d Hubbard
model\cite{PhysRevB.6.930,schulz1991correlated}.  In the large $U$
limit, we obtain $W \sim 2\pi^2 t^2/(3U)$, while for small $U$, $W\sim
U$.   Using the former estimate, we obtain $W \approx (1-2)t$ for the titanates.

Consequently we can obtain the free energy for an {\em arbitrary}
magnetization of the 1d $xz$ and $yz$ subbands. If the spin (per site) in the $xy$ band is $M_0$ and that in
the $xz$ and $yz$ bands is $M_1$, it is (per site)
\begin{equation}
  \label{eq:4}
  F = \frac{M_0^2 }{2\chi_{2d}} + 2 x_1 J_{\rm eff} F_{1}\left[
    \frac{M_1}{x_1}, \frac{k_B T}{J_{\rm eff}}\right] -  J_H (2 M_0 M_1 + M_1^2),
\end{equation}
where $F_{1}[m,t]$ is the free energy per site of the 1d
antiferromagnet chain unit exchange with magnetization $m$ and
temperature $t$.  This assumes $x_1 \ll 1$.  The first two terms
represent the exact thermodynamics for the decoupled orbital
subsystems, and fully incorporate all the effects of $U$.  The last
term is simply the leading first order term in the expansion of the
energy of these states in $J_H$, presumed small.  Eq.~\eqref{eq:4} may
also be interpreted in terms of a mean-field treatment {\em of the
  Hund's coupling only}.  This is quite analogous to ``chain mean
field theory'', which has been successfully applied to explain
numerous experiments in low dimensional magnetic materials,\cite{PhysRevB.82.014421} and is
known to be usually quantitatively rather accurate.

Using $J_{\rm eff} = W x_1^3$ and $M_1 \leq x_1/2$, we see from
Eq.~\eqref{eq:4} that when
$x_1 \lesssim \sqrt{J_H/W}$, the Hund's energy overwhelms the 1d
exchange and favors a ferromagnetic state with $M_1 \neq 0$.
Remarkably, this occurs for {\sl arbitrarily weak} Hund's coupling
$J_H$, provided the filling of the upper $xz$ and $yz$ subbands is
sufficiently small, and of course non-zero.
This gives a mechanism for ferromagnetism at {\sl intermediate carrier
  density}, when the total density is near the critical value needed
to just populate the $xz$ and $yz$ subbands, with magnetism
disappearing both for smaller and larger carrier density.  A
quantitative minimization of Eq.~\eqref{eq:4} is possible since $F_1[m,t]$
is known exactly from the thermodynamic Bethe
ansatz\cite{klumper1998spin}.    Assuming $J_H \ll t, W$ and $x_1 \ll
1$, we obtain a dome-shaped region of ferromagnetism, as shown in
Fig.~\ref{fig:phase}.  Note that the characteristic maximum temperature scale is of
order of $k_B T_c \sim 0.05\sqrt{J_H^3/W}$.  From this minimization we
can also obtain the magnetization at all temperatures and fields.  In
particular we find that at $T=0$ the $xz$/$yz$ bands are
fully polarized.

It may appear that the one dimensional physics of the above picture is
overly exotic and restrictive.  However, this is not the case, and can
persist up to some reasonable value of $t'$.
This hopping causes a crossover from 1d behavior to 2d
Fermi liquid behavior at low energy.  By continuity, for small $t'/t$,
this Fermi liquid must have an enhanced spin susceptibility captured
by a large Fermi liquid correction $F_0^a$.  However, the eventual
two-dimensionality induced by non-zero $t'$ controls the maximum
susceptibility achieved at
small $x_1$, and if this effect is too large, the ferromagnetic
instability may be entirely removed.  The susceptibility divergence is cut off when
the distance between the Fermi energy and the bottom of the $xz$ and
$yz$ bands is comparable to the hopping $t'$, i.e. $t x_1^2 \sim t'$.
The same condition describes the change from an open Fermi surface to
an elliptical one.  This gives the condition $t' \lesssim J_H t/W$ for
the ferromagnetic phase to occur (we neglect numerical prefactors here
due to the imprecision of the matching argument).  Of course, when
$t'$ is substantial, the magnitude of the magnetization and of $T_c$ will be
reduced from the 1d values given in Fig.~\ref{fig:phase}, further
increasing the tendency to low $T_c$ and small net moment.

\begin{table}[htp]
\centering
\begin{tabular}{c|c|c}
    \hline
Interfaces & Symmetry &   Local orbitals      \\ \hline
    (001)  &  4-fold rotation&   $xz$, $yz$; $xy $     \\     \hline
    (110)   &   2-fold rotation &  $\frac{1}{\sqrt{2}}(xz+yz)$;  $xy$;  $\frac{1}{\sqrt{2}}(xz-yz)$ \\ \hline
              &       &     $  \frac{1}{\sqrt{3}} (xy+e^{i \frac{2\pi}{3} }yz+e^{-i \frac{2\pi}{3} }xz)$, \\
   (111)  & 3-fold rotation &  $\frac{1}{\sqrt{3}} (xy+e^{-i \frac{2\pi}{3} }yz+e^{i \frac{2\pi}{3} }xz)$;  
 \\
& &  $ \frac{1}{\sqrt{3}} (xy+yz+xz)$ \\
    \hline
\end{tabular}
\caption{The relevant local crystal symmetries and local orbital states for different interfaces. 
`;' delimits the sets of locally degenerate orbital states. 
Note that there could be a small hybridization between $\frac{1}{\sqrt{2}}(xz+yz)$ and $xy$ orbitals for the (110) interface. }
\label{tab:interfaces}
\end{table}

Recently the
(110) and (111) LAO-STO interfaces have also been prepared
experimentally\cite{Herranz12, Annadi12}, and both interfaces appear to
support STO electron gases, though the (111) interface is polar and the
(110) is not.  As listed in Tab.~\ref{tab:interfaces}, these two
interfaces have different local crystal field environments and hence
different local orbital configuration from the (001) interface.  Can
these two interfaces also support ferromagnetism -- under ideal
disorder-free conditions -- at certain electron
fillings?

As usual, the e$_g$ doublets are always higher in energy and do not play
any role.  For the (110) interface, the three t$_{2g}$ orbitals are
splitted into three non-degenerate orbitals,
$\frac{1}{\sqrt{2}}(xz+yz)$, $\frac{1}{\sqrt{2}}(xz-yz)$ and $xy$.  In the
first approximation, the local hybridization between
$\frac{1}{\sqrt{2}}(xz+yz)$ and $xy$ orbitals may be neglected. When
these two orbitals form bands, they are also
quasi-one-dimensional just like $xz$ and $yz$ orbitals for the (001)
interface. Hopping among $\frac{1}{\sqrt{2}}(xz+yz)$ (/$xy$) orbitals
occurs most strongly with neighbors along $z$ ( /[1$\bar{1}$0]) lattice
directions. The $\frac{1}{\sqrt{2}}(xz-yz)$ subband is two dimensional
and its band bottom is the lowest among the three subbands.  Due to the
reduced symmetry of the (110) interface, the two upper
quasi-one-dimensional subbands are split. Based on our above
discussion of ferromagnetic instability for the (001) interface, we also
expect emergent ferromagnetism for the (110) interface when the filling
of the quasi-one-dimensional subband is sufficiently small. Because the
two upper quasi-one-dimensional subbands are not degenerate, there may
even exist two ferromagnetic regime as the electron filling of the two
upper subband is increased.  One should note that the discussion here
assumes no hybridization between $\frac{1}{\sqrt{2}}(xz+yz)$ and $xy$
orbitals. In reality, there are always small hybridization between these
two orbitals. If this hybridization is very small (smaller than
$\mathcal{O} (\sqrt{ J_H/J } )$), the resulting two-dimensional Fermi
liquid should still have a large spin susceptibility and ferromagnetism
can still be present.

For the (111) interface, although locally the crystal field splits three t$_{2g}$ orbitals into 
one a$_{1g}$ state, $\frac{1}{\sqrt{3}} (xy+yz+xz)$, and two e$_{2g}'$
states, $\frac{1}{\sqrt{3}} (xy+e^{i \frac{2\pi}{3} }yz+e^{-i \frac{2\pi}{3} }xz)$ and 
$\frac{1}{\sqrt{3}} (xy+e^{-i \frac{2\pi}{3} }yz+e^{i \frac{2\pi}{3}
}xz)$, the electron hopping strongly hybridizes three orbitals and
leads to two-dimensional Fermi liquids. Hence no ferromagnetism arises
in this case.

In constrast to the itinerant mechanism discussed here,  other theoretical works have instead
proposed mechanisms relying on localized electron moments.  While we
believe that {\em Mott} localization of electrons near the interface
should not occur for ideal structures, sufficient disorder and
interactions together might create some truly localized moments.  If
the localized electron mechanisms are correct, we predict significant
dependence of the ferromagnetism on disorder, and indeed that it
should weaken as sample quality is improved.  The itinerant mechanism
discussed here has its own distinct predictions,
e.g. Fig.~\ref{fig:phase}, and the fact that the polarization resides
in $xz$/$yz$ bands, which may be tested by x-ray dichroism experiments. Further, varying the electron concentration
away from the critical density by tuning the back gate voltage may
easily suppress the ferromagnetism, and ferromagnetism should be
absent at the (111) interface; neither prediction applies for the
local moment mechanism\cite{PhysRevLett.108.117003}.

We thank Jim Allen, Lu Li and Susanne Stemmer for helpful discussions.
GC was supported by DOE award no. DE-SC0003910. 
LB was supported by DARPA through Grant No. W911-NF-12-1-0574 . 
Some of this work was carried out at the Aspen Center for Physics and the Kavli Institute for Theoretical Physics; our stays there were supported in part by NSF grant no. 1066293, and NSF grant no. PHY11-25915, respectively.

\bibliography{refs}

\end{document}